\documentstyle[preprint,aps,epsf]{revtex}  

\catcode`@=11
\def\chkspace{%
  \relax   
  \begingroup\ifhmode\aftergroup\dochksp@ce\fi\endgroup}
\def\dochksp@ce{%
  \unskip              
  \futurelet\chkspct@k\d@chkspc  
}
\def\d@chkspc{%
  \let\nxtsp@ce=\relax
  \ifx\chkspct@k.\else     
    \ifx\chkspct@k,\else
      \ifx\chkspct@k;\else
        \ifx\chkspct@k!\else
          \ifx\chkspct@k?\else
            \ifx\chkspct@k:\else
              \ifx\chkspct@k)\else
              \ifx\chkspct@k(\else
                \ifx\chkspct@k]\else
                  \ifx\chkspct@k-\else
                    \ifx\chkspct@k\egroup\else  
                      \let\nxtsp@ce=\put@space  
                    \fi
                  \fi
                \fi
              \fi
              \fi
            \fi
          \fi
        \fi
      \fi
    \fi
  \fi
  \nxtsp@ce
}
\def\put@space{$\;$}
\catcode`@=12

\def\z0{{$Z^0$}\chkspace}
\def\Dst{\relax\ifmmode {\rm D}^* \else {D$^*$}\fi\chkspace}
\def\Dpl{\relax\ifmmode {\rm D}^+ \else {D$^+$}\fi\chkspace}
\def\D0{\relax\ifmmode {\rm D}^0 \else {D$^0$}\fi\chkspace}
\def\Kst{\relax\ifmmode {\rm K}^* \else {K$^*$}\fi\chkspace}
\def\K0{\relax\ifmmode {\rm K}^0_s \else {K$^0_s$}\fi\chkspace}
\def\Kpl{\relax\ifmmode {\rm K}^+ \else {K$^+$}\fi\chkspace}
\def\Kstz{\relax\ifmmode {\rm K}^{*0} \else {K$^{*0}$}\fi\chkspace}

\def\ra{{$\rightarrow$}\chkspace}

\def\be{\begin{equation}}
\def\ee{\end{equation}}
\def\bea{\begin{eqnarray}}
\def\eea{\end{eqnarray}}

\def\ep{{$e^+e^-$}\chkspace}

\def\z0{$Z^0$}

\def\bb{{$b\bar{b}$}\chkspace}
\def\cc{{$c\bar{c}$}\chkspace}

\def\etal{{\it et al.}\chkspace}
\def\ie{{\it i.e.}\chkspace}
\def\eg{{\it e.g.}\chkspace}
\def\adhoc{{\it ad hoc}\chkspace}

\begin{document}


\thispagestyle{empty}
\begin{flushright}
{\renewcommand{\baselinestretch}{.75}
  SLAC--PUB--8316\\
December 1999\\
}
\end{flushright}

\vskip 1truecm
 

\begin{center}
{\large\bf
Precise Measurement of the $b$-quark Fragmentation  
Function in Z$^{0}$ Boson Decays$^*$
}
\end{center}
 
\vspace {1.0cm}
 
\begin{center}
 {\bf The SLD Collaboration$^{**}$}\\
Stanford Linear Accelerator Center \\
Stanford University, Stanford, CA~94309
\end{center}
 
\vspace{1cm}

\begin{abstract}
\noindent
We have developed a new technique for inclusive reconstruction of the
energy of $B$ hadrons. The excellent efficiency and resolution 
of this technique allow us to make the most precise
determination of the $b$-quark fragmentation function, using \ep \ra \z0
decays recorded in the SLD experiment at SLAC. We compared our
measurement with the predictions of a number of fragmentation models.
We excluded several of these models and
measured the average scaled energy 
of weakly-decaying $B$ hadrons to be $\left<x_B\right>$ $=$ 0.714 
$\pm$ 0.005 (stat) $\pm$ 0.007 (syst) $\pm$ 0.002 (model-dependence). 

\end{abstract}

\vskip 1truecm
\centerline{\it (Submitted to Physical Review Letters)}

\vskip 1truecm



\vfill
\noindent
{\footnotesize
$^*$ Work supported by Department of Energy contract DE-AC03-76SF00515 (SLAC).}

\eject

In high-energy strong-interaction processes quarks and gluons are not observed 
directly, but appear as jets of colorless hadrons.
This fundamental process of `jet fragmentation' affects all high-energy
physics measurements involving strongly-interacting particles, but is
only poorly understood at a quantitative level.
The fragmentation of heavy 
quarks is of particular experimental interest since many expected signatures of
new heavy particles, such as Higgs and SUSY particles, involve decays to 
$b$ quarks. It is hence vital to understand the production and properties of
$b$ jets. Here we present a significantly improved determination of the
$b$-quark fragmentation function, $D(x_B)=(1/\sigma)d\sigma/dx_B$, measured using
\z0 \ra \bb decays, where $x_B$ = $E_B/E_{beam}$ represents the fraction of
the $b$-quark energy retained by the weakly-decaying $B$ hadron.

In Quantum Chromodynamics (QCD) the $b$-quark mass serves as a cutoff for
collinear gluon radiation.
The distribution of $b$-quark energies prior to hadronisation can therefore be 
calculated perturbatively~\cite{mn,jaffe,lisa,bcfy,dkt}. 
However, the additional  
effects which yield the experimentally-accessible distribution $D(x_B)$
are non-perturbative, and have been studied phenomenologically via 
a number of different 
approaches~\cite{mn,jaffe,lisa,bcfy,dkt,cn,webber,kart,bowler,pete,lund,collins}.
Measurements of $D(x_B)$ provide direct tests of these perturbative QCD and
model predictions.

$D(x_B)$ has been measured previously~\cite{delphi93,aleph95,opal95,sld96} 
by reconstructing the energies of $B$ hadrons that decay semi-leptonically
($B$ \ra $l\nu DX$) in \z0 \ra \bb events.
In these studies the inclusive $B$ selection efficiency was smaller than 1\%,
and much lower for $E_B$ $<$ 20 GeV.
The resulting low-statistics samples, and limited energy resolutions,
yielded poor constraints on the shape of the distribution.

We have developed a new technique for measuring $E_B$ using only 
kinematic information from charged tracks.
Our 307 Mpixel CCD-based vertex detector, combined with the micron-sized
SLC interaction point (IP), allows us to reconstruct $B$-decay vertices
and the $B$-hadron flight direction very accurately. 
The method yields a significantly higher $B$-selection efficiency and superior energy 
resolution, 
both of which are almost independent of $E_B$, and has very low bias since it
does not use a beam-energy constraint.
This allows us to measure the shape of $D(x_B)$ with sufficient precision
to make stringent tests of $b$-fragmentation model predictions, 
reduce model-dependent systematic errors, and therefore
discriminate among these models for the first time.
Furthermore, our technique for
reconstructing $B$-hadron energies is of direct relevance to studies of
other important properties of heavy-quark systems,
such as $B$-hadron lifetimes and neutral $B$-meson mixing.

We used 150,000 hadronic \z0 decays produced in \ep annihilations at the
SLAC Linear Collider (SLC) and collected in the 
SLC Large Detector (SLD) between 1996 and 1997.
A description of the SLD can be found 
elsewhere~\cite{sld1,sld2}. The trigger and selection criteria
for \z0 \ra hadrons events are described elsewhere~\cite{sld96}.
This analysis used charged tracks measured in the Central Drift
Chamber (CDC)~\cite{cdc}
and in the upgraded CCD Vertex Detector (VXD)~\cite{vxd3}, with a
momentum resolution of
$\sigma_{p_{\perp}}/p_{\perp}$ = $0.01\oplus0.0026p_{\perp}$,
where $p_{\perp}$ is the track transverse momentum with respect
to the beamline, in GeV/$c$. The centroid of the SLC IP 
was reconstructed with a precision of 
4.4$\mu$m (30$\mu$m) in the plane transverse to (containing) the beamline.
Tracks from identified $\gamma$ conversions and $K^0$ or $\Lambda^0$ decays
were removed from consideration, and only well-reconstructed
tracks~\cite{sld2} were used for $B$-hadron tagging
and energy reconstruction. 

Weakly-decaying $B$ hadrons were identified by exploiting their 
long lifetimes and large masses relative to light-flavor hadrons.
Each hadronic event was divided into two hemispheres by the plane
containing the IP and the normal to the thrust axis.
A topological vertexing algorithm~\cite{zvnim} 
was optimized for this analysis and applied to the set of well-reconstructed
tracks in each 
hemisphere in an attempt to reconstruct a $B$-decay vertex.
Vertices were required to be separated from the
IP by at least 1 mm and to contain at least two tracks.
A candidate vertex was found in 32,492 hemispheres~\cite{twobs}.

In each hemisphere the total energy, $E_{ch}$, momentum, $\vec{P_{ch}}$, and
invariant mass, $M_{ch}$, of the vertex-associated tracks
were calculated by assigning each track the charged-pion mass.  
Due to neutral decay products and tracks missed from the vertex, $\vec{P_{ch}}$
can be acollinear with the true $B$ flight direction, which we estimated
independently of the track momenta by the unit vector $\vec{v}$ along the
line joining the IP to the reconstructed vertex position.
Our accurate knowledge of the IP and vertex positions yielded
an excellent angular resolution of 26 mrad on $\vec{v}$.
This allowed us to calculate the net transverse momentum,
relative to $\vec{v}$, of the particles from the $B$ decay
that were not associated with the vertex, $\vec{P_t}=\vec{P_{ch}}
 - (\vec{P_{ch}} \cdot \vec{v})\,\vec{v}$,
and hence to improve our estimate of the mass of the decaying particle via
$M_{P_t} \equiv \sqrt{M_{ch}^2 + P_t^2} + |P_t|$; this quantity is a strong
discriminator for selecting $B$ hadrons~\cite{sldrb}.  
We required 2.0 GeV/$c^{2}<M_{P_t}<2\times M_{ch}$
to select 19,604 candidates, estimated 
to be 98.2\% $B$ hadrons, the main background (1.6\%) being 
charmed hadrons in \z0 \ra \cc events.
The efficiency for selecting a true $B$ hadron is 40.1\%. 
For this sample, on average 92\% of the reconstructed 
true $B$-decay tracks were associated with the  
vertex, and 98\% of the vertex-associated tracks were from true 
$B$ decays. 

The reconstructed $B$-hadron energy, $E_B^{rec}$, can be expressed as
$\sqrt{M_0^2+P_t^2+P_{0l}^2}+E_{ch}$, where the combined mass of the missing 
particles, $M_0$, and the missing momentum along $\vec{v}$, $P_{0l}$,
are the only unmeasured quantities.
If we assume a $B$-hadron rest mass, $M_B$, we can eliminate one of the two
unknowns and calculate an upper bound on $M_0$:
\begin{equation}
M_{0}^2 \leq M_B^2 - 2M_B\sqrt{M_{ch}^2 + P_t^2} + M_{ch}^2 \equiv M_{0max}^2,
\end{equation}
where equality holds when $P_{0l}$ = 0 in the $B$-hadron rest frame.
Since, from phase-space requirements, small values of $P_{0l}$ are the most probable,  
and the average $B$-decay multiplicity is high, the true $M_0$ tends
to be close to $M_{0max}$~\cite{ddthesis}.
We assumed $M_B$ = 5.28 GeV/$c^2$,
equated $M_0$ with the
measured $M_{0max}$, and calculated $E_B^{rec}$.
This estimate of the $B$ energy is best when the $B$-decay kinematics are 
well constrained by the vertex-associated tracks, \ie when
$M_{0max}$ $\simeq$ 0~\cite{negm}.
Also, the small non-$B$ background is concentrated at large $M_{0max}$.
In order to improve the energy resolution, and reduce further the background,
we required $-1 < M_{0max}^2 < [1.1+f(E_B^{rec})]^2$, where the
explicit $E_B^{rec}$-dependence \cite{edep} was chosen so that the $B$ selection
efficiency is only weakly energy-dependent. The efficiency is
above 3\% for $E_B$ $>$ 10 GeV; the average value is 3.9\%.
1920 candidates were selected, with an estimated $B$ purity of 99.3\%.
 
We examined the distribution of the normalized difference 
between the reconstructed and true 
$B$-hadron energies, $(E_B^{rec}-E_B^{true})/E_B^{true}$. 
This resolution can be characterised~\cite{ddthesis} by a double Gaussian
function centered at zero.
We found that the narrower Gaussian represents 83\% of the population
and has a width of 10.4\%.
This resolution depends only weakly on $E_B^{true}$; in particular
it remains better than 15\% even for $B$ energies close to the mass threshold, 
which is a significant advantage of this technique.
The estimated 0.7\% non-$B$ background was subtracted bin-by-bin
from the distribution of the reconstructed scaled 
$B$-hadron energy, $D^{rec}(x_B^{rec})$, which is shown in Fig.~1.

We tested several models of $b$-quark fragmentation.
These models are formulated in terms of experimentally inaccessible
variables and must hence be implemented in an iterative fragmentation algorithm
in order to derive the measurable quantity $D^{rec}(x_B^{rec})$. 
We employed our JETSET 7.4~\cite{jetset}-based Monte Carlo simulation 
program~\cite{sld96,sldrb} 
to generate \ep \ra \bb events according to each model considered.  
$B$-hadron energies were reconstructed, 
according to our algorithm, from the fully-simulated event sample
to derive $D^{sim}(x_B^{rec})$, which was compared with the data using a binned
$\chi^2$. 
The $\chi^2$ was minimised by repeating this procedure under
variation of the input parameter(s) of the model.
The fitted model predictions and best $\chi^2$ values are shown in Fig.~1. 

Within this context the 
Bowler~\cite{bowler,jetset}, Kartvelishvili \etal~\cite{kart} and 
Lund~\cite{lund} models reproduce the data;
the models of Braaten \etal ~\cite{bcfy}, Collins and Spiller~\cite{collins}
and Peterson \etal~\cite{pete} 
have a $\chi^2$ confidence level less than 0.1\% 
and are not consistent with the data. 
We also tested the Monte Carlo models HERWIG 5.7~\cite{herwig} and
UCLA~\cite{ucla}, which contain no explicit free parameters to
control $D(x_B)$. The UCLA model is consistent with the data and the
HERWIG model is not (Fig.~1). 

In order to allow other models to be compared with our data,
\eg those of~\cite{mn,jaffe,lisa,dkt,cn,webber}, we corrected for the effects of
the selection and reconstruction procedures that were applied. 
We estimated the {\it true} weakly-decaying $B$-hadron scaled-energy
distribution, $D(x_B)$, from the (background-subtracted) reconstructed
distribution, $D^{rec}(x_B^{rec})$;  for each bin $i$:
$D(x_B)_i = \Sigma_k M_{ik} D^{rec}(x_B^{rec})_k / \epsilon_i$.
The selection efficiency $\vec{\epsilon}$ and the unfolding matrix $\bf M$ were
calculated from the simulation using in turn each of the four fragmentation
models that were consistent with the data, with the respective fitted parameter
values.
This unfolding procedure assumes a smooth underlying distribution and is
explicitly model dependent, but we quantify this (below) using the variation 
among the four resulting distribution shapes.

This set of consistent models is small and does not appear to span the range of potentially
acceptable shapes.
We therefore considered a number of \adhoc functional forms for the 
true $x_B$ distribution, and found four that yielded a reconstructed
distribution consistent with the data in Fig.~1: an
8th-order polynomial, the Peterson function,
and two generalisations of the Peterson function~\cite{aleph95}. The 
respective parameter value(s) was (were) optimised in a manner similar to that
described above.
Each of these four fitted functions was also used to
calculate $\vec{\epsilon}$ and $\bf M$ from the simulation and to derive $D(x_B)$.

Each of the eight unfolded distributions was normalized to unit area, and
in Fig.~2 we show their binwise average;
the band includes the r.m.s. deviation. This represents our
best estimate of the true $B$ energy distribution $D(x_B)$.
The corrected distribution is, by construction, smoother than the measured
distribution, and the band provides an envelope within which  
acceptable predictions should fall.  
The constraint on the shape is much stronger than in any previous
determination~\cite{delphi93,aleph95,opal95,sld96}. 

An important advantage of our
method is that it is not biased by tracks that were not reconstructed 
or attached to the vertex~\cite{ddthesis}.  
We considered other potential sources of systematic uncertainty 
which derive from the modelling of our detector response.  
In each case the simulated events were reweighted or adjusted in order to
reproduce the variation in question, and the entire analysis was repeated. 
Variation of quantities such as the track reconstruction efficiency,
the point resolution of the vertex detector and the momentum and dip angle
resolutions of the tracking system within their uncertainties~\cite{sldrb}
affects the $B$ selection efficiency and, in some cases, the energy resolution.
However, there is little effect on the shape of the distribution or on the
mean value, $\left<x_B\right>$.  
In no case was the $\chi^2$ of a model or function test changed significantly.  
The largest effect on $\left<x_B\right>$, $\pm$0.005, arose from the uncertainty in
the momentum resolution, which was measured in the data
using \ep \ra $\mu^+\mu^-$ events.  
An \adhoc correction was 
applied to the simulation to reproduce the measurement, and 
the full effect of this correction was considered as a symmetric systematic
uncertainty. As a cross check we
changed the upper cut on $M_{0max}^2$ to fixed values between  
1 and 5 (GeV/$c^2$)$^2$; the change in $\left<x_B\right>$
was smaller than the statistical error.
 
We also considered the uncertainties on a large number of measured 
quantities~\cite{sldrb} related to the 
production and decay of $B$ and charmed hadrons. These are used in the
simulation
and potentially affect the values of $\vec{\epsilon}$ and {\bf M} used to
unfold the data. We varied each quantity by the error on
its measured value; none of these variations affects our
conclusions. The production of primary excited $B$ hadrons, 
collectively denoted $B^*$ and $B^{**}$, which decay into the weakly-decaying
$B$ hadrons that we measure, was investigated in more detail, and we varied 
independently the
fraction of primary $B^*$ ($B^{**}$) from zero to unity (0.5).  No significant
effect on the shape of $D(x_B)$, or on the $\chi^2$ values for the model
comparisons, was observed.
The largest effect on $\left<x_B\right>$, of $\pm0.002$, was due to variation
of the number of charged tracks and K$^0_s$ produced per $B$-hadron decay.
In each $x_B$ bin the sum in quadrature of the detector- and physics-related 
systematic uncertainties on $D(x_B)$ is much smaller than the statistical
error;  they are not shown in Fig.~2.  

From the eight distribution shapes that are consistent with our data 
we extract the mean value of the scaled 
energy of weakly-decaying $B$ hadrons in $Z^0$ decays:
$$
\left< x_B \right> = 0.714 \pm 0.005({\rm stat.}) \pm 0.007 ({\rm syst.}) 
\pm 0.002({\rm model}).
$$
This is the most precise of the world's measurements that take the $D(x_B)$
shape dependence into account, and the uncertainty is relatively small since 
we have excluded a wide range of shapes.
Our result is consistent with a recent average over many \z0 measurements
of  
$\left< x_B \right> = 0.702\pm0.008$~\cite{lephf}.
We also calculated the second and third moments to be
$\left< x_B^2 \right> = 0.537 \pm 0.011$ (exp.) $\pm 0.003$ (model) and
$\left< x_B^3 \right> = 0.417 \pm 0.012$ (exp.) $\pm 0.004$ (model).

In order to derive results for the inclusive sample of primary $B$ hadrons, 
one must assume 
values for the production fractions of $B^{*}$ and $B^{**}$ mesons.
Postulating a $B^{*}$ production fraction of 0.75, expected from naive spin
counting, leads to $\left<x_B\right>^{primary}$ = 0.718.
Postulating independently a $B^{**}$ production fraction of
0.25
yields $\left<x_B\right>^{primary}$ = 0.728.

In summary, we have developed a new, inclusive technique for 
reconstructing the energies of $B$ hadrons.
It has substantially higher efficiency and better energy resolution than previous
methods.
We have employed this technique to measure the scaled-energy distribution 
of weakly-decaying $B$ hadrons produced in $e^+e^-$ \ra \z0 decays with
unprecedented precision over the entire kinematic range from the $B$
mass to the beam energy.
As a result we are able to exclude several models of $b$-quark
fragmentation, including the widely-used JETSET$+$Peterson model.

\vskip 1truecm

We thank the personnel of the SLAC accelerator department and the
technical
staffs of our collaborating institutions for their outstanding efforts
on our behalf. 
We thank M.~Bowler and M.~Seymour for helpful comments.
This work was supported by the U.S. Department of Energy, 
  the UK Particle Physics and Astronomy Research Council
  (Brunel, Oxford and RAL);
  the Istituto Nazionale di Fisica Nucleare of Italy
  (Bologna, Ferrara, Frascati, Padova, Perugia, Pisa);
  the Japan-US Cooperative Research Project on High Energy Physics
  (Nagoya, Tohoku);
   and the Korea Science and Engineering Foundation (Soongsil).

\vfil
\eject

\noindent
{\bf $^{**}$List of Authors}

%
%
\begin{center}
\def\iAOMORI{$^{(1)}$}
\def\iBRI{$^{(2)}$}
\def\iBRUN{$^{(3)}$}
\def\iBU{$^{(4)}$}
\def\iCOLO{$^{(5)}$}
\def\iCSU{$^{(6)}$}
\def\iFERR{$^{(7)}$}
\def\iFRAS{$^{(8)}$}
\def\iJHU{$^{(9)}$}
\def\iLBL{$^{(10}$}
\def\iMASS{$^{(11)}$}
\def\iMISSI{$^{(12)}$}
\def\iMIT{$^{(13)}$}
\def\iMOSCOW{$^{(14)}$}
\def\iNAGO{$^{(15)}$}
\def\iOREG{$^{(16)}$}
\def\iOXF{$^{(17)}$}
\def\iPERU{$^{(18)}$}
\def\iRAL{$^{(19)}$}
\def\iRUTG{$^{(20)}$}
\def\iSLAC{$^{(21)}$}
\def\iSOONG{$^{(22)}$}
\def\iTENN{$^{(23)}$}
\def\iTOHO{$^{(24)}$}
\def\iUCSB{$^{(25)}$}
\def\iUCSC{$^{(26)}$}
\def\iVAND{$^{(27)}$}
\def\iWASH{$^{(28)}$}
\def\iWISC{$^{(29)}$}
\def\iYALE{$^{(30)}$}

\baselineskip=.75\baselineskip 

\mbox{Kenji  Abe\unskip,\iNAGO}
\mbox{Koya Abe\unskip,\iTOHO}
\mbox{T. Abe\unskip,\iSLAC}
\mbox{I. Adam\unskip,\iSLAC}
\mbox{H. Akimoto\unskip,\iSLAC}
\mbox{D. Aston\unskip,\iSLAC}
\mbox{K.G. Baird\unskip,\iMASS}
\mbox{C. Baltay\unskip,\iYALE}
\mbox{H.R. Band\unskip,\iWISC}
\mbox{T.L. Barklow\unskip,\iSLAC}
\mbox{J.M. Bauer\unskip,\iMISSI}
\mbox{G. Bellodi\unskip,\iOXF}
\mbox{R. Berger\unskip,\iSLAC}
\mbox{G. Blaylock\unskip,\iMASS}
\mbox{J.R. Bogart\unskip,\iSLAC}
\mbox{G.R. Bower\unskip,\iSLAC}
\mbox{J.E. Brau\unskip,\iOREG}
\mbox{M. Breidenbach\unskip,\iSLAC}
\mbox{W.M. Bugg\unskip,\iTENN}
\mbox{D. Burke\unskip,\iSLAC}
\mbox{T.H. Burnett\unskip,\iWASH}
\mbox{P.N. Burrows\unskip,\iOXF}
\mbox{A. Calcaterra\unskip,\iFRAS}
\mbox{R. Cassell\unskip,\iSLAC}
\mbox{A. Chou\unskip,\iSLAC}
\mbox{H.O. Cohn\unskip,\iTENN}
\mbox{J.A. Coller\unskip,\iBU}
\mbox{M.R. Convery\unskip,\iSLAC}
\mbox{V. Cook\unskip,\iWASH}
\mbox{R.F. Cowan\unskip,\iMIT}
\mbox{G. Crawford\unskip,\iSLAC}
\mbox{C.J.S. Damerell\unskip,\iRAL}
\mbox{M. Daoudi\unskip,\iSLAC}
\mbox{S. Dasu\unskip,\iWISC}
\mbox{N. de Groot\unskip,\iBRI}
\mbox{R. de Sangro\unskip,\iFRAS}
\mbox{D.N. Dong\unskip,\iMIT}
\mbox{M. Doser\unskip,\iSLAC}
\mbox{R. Dubois\unskip,\iSLAC}
\mbox{I.Erofeeva\unskip,\iMOSCOW}
\mbox{V. Eschenburg\unskip,\iMISSI}
\mbox{E. Etzion\unskip,\iWISC}
\mbox{S. Fahey\unskip,\iCOLO}
\mbox{D. Falciai\unskip,\iFRAS}
\mbox{J.P. Fernandez\unskip,\iUCSC}
\mbox{K. Flood\unskip,\iMASS}
\mbox{R. Frey\unskip,\iOREG}
\mbox{E.L. Hart\unskip,\iTENN}
\mbox{K. Hasuko\unskip,\iTOHO}
\mbox{S.S. Hertzbach\unskip,\iMASS}
\mbox{M.E. Huffer\unskip,\iSLAC}
\mbox{X. Huynh\unskip,\iSLAC}
\mbox{M. Iwasaki\unskip,\iOREG}
\mbox{D.J. Jackson\unskip,\iRAL}
\mbox{P. Jacques\unskip,\iRUTG}
\mbox{J.A. Jaros\unskip,\iSLAC}
\mbox{Z.Y. Jiang\unskip,\iSLAC}
\mbox{A.S. Johnson\unskip,\iSLAC}
\mbox{J.R. Johnson\unskip,\iWISC}
\mbox{R. Kajikawa\unskip,\iNAGO}
\mbox{M. Kalelkar\unskip,\iRUTG}
\mbox{R.R. Kofler\unskip,\iMASS}
\mbox{R.S. Kroeger\unskip,\iMISSI}
\mbox{M. Langston\unskip,\iOREG}
\mbox{D.W.G. Leith\unskip,\iSLAC}
\mbox{V. Lia\unskip,\iMIT}
\mbox{C.Lin\unskip,\iMASS}
\mbox{G. Mancinelli\unskip,\iRUTG}
\mbox{S. Manly\unskip,\iYALE}
\mbox{G. Mantovani\unskip,\iPERU}
\mbox{T.W. Markiewicz\unskip,\iSLAC}
\mbox{T. Maruyama\unskip,\iSLAC}
\mbox{A.K. McKemey\unskip,\iBRUN}
\mbox{R. Messner\unskip,\iSLAC}
\mbox{K.C. Moffeit\unskip,\iSLAC}
\mbox{T.B. Moore\unskip,\iYALE}
\mbox{M.Morii\unskip,\iSLAC}
\mbox{D. Muller\unskip,\iSLAC}
\mbox{V. Murzin\unskip,\iMOSCOW}
\mbox{S. Narita\unskip,\iTOHO}
\mbox{U. Nauenberg\unskip,\iCOLO}
\mbox{G. Nesom\unskip,\iOXF}
\mbox{N. Oishi\unskip,\iNAGO}
\mbox{D. Onoprienko\unskip,\iTENN}
\mbox{L.S. Osborne\unskip,\iMIT}
\mbox{R.S. Panvini\unskip,\iVAND}
\mbox{C.H. Park\unskip,\iSOONG}
\mbox{I. Peruzzi\unskip,\iFRAS}
\mbox{M. Piccolo\unskip,\iFRAS}
\mbox{L. Piemontese\unskip,\iFERR}
\mbox{R.J. Plano\unskip,\iRUTG}
\mbox{R. Prepost\unskip,\iWISC}
\mbox{C.Y. Prescott\unskip,\iSLAC}
\mbox{B.N. Ratcliff\unskip,\iSLAC}
\mbox{J. Reidy\unskip,\iMISSI}
\mbox{P.L. Reinertsen\unskip,\iUCSC}
\mbox{L.S. Rochester\unskip,\iSLAC}
\mbox{P.C. Rowson\unskip,\iSLAC}
\mbox{J.J. Russell\unskip,\iSLAC}
\mbox{O.H. Saxton\unskip,\iSLAC}
\mbox{T. Schalk\unskip,\iUCSC}
\mbox{B.A. Schumm\unskip,\iUCSC}
\mbox{J. Schwiening\unskip,\iSLAC}
\mbox{V.V. Serbo\unskip,\iSLAC}
\mbox{G. Shapiro\unskip,\iLBL}
\mbox{N.B. Sinev\unskip,\iOREG}
\mbox{J.A. Snyder\unskip,\iYALE}
\mbox{H. Staengle\unskip,\iCSU}
\mbox{A. Stahl\unskip,\iSLAC}
\mbox{P. Stamer\unskip,\iRUTG}
\mbox{H. Steiner\unskip,\iLBL}
\mbox{D. Su\unskip,\iSLAC}
\mbox{F. Suekane\unskip,\iTOHO}
\mbox{A. Sugiyama\unskip,\iNAGO}
\mbox{M. Swartz\unskip,\iJHU}
\mbox{F.E. Taylor\unskip,\iMIT}
\mbox{J. Thom\unskip,\iSLAC}
\mbox{T. Usher\unskip,\iSLAC}
\mbox{J. Va'vra\unskip,\iSLAC}
\mbox{R. Verdier\unskip,\iMIT}
\mbox{D.L. Wagner\unskip,\iCOLO}
\mbox{A.P. Waite\unskip,\iSLAC}
\mbox{S. Walston\unskip,\iOREG}
\mbox{J. Wang\unskip,\iSLAC}
\mbox{A.W. Weidemann\unskip,\iTENN}
\mbox{E. R. Weiss\unskip,\iWASH}
\mbox{J.S. Whitaker\unskip,\iBU}
\mbox{S.H. Williams\unskip,\iSLAC}
\mbox{S. Willocq\unskip,\iMASS}
\mbox{R.J. Wilson\unskip,\iCSU}
\mbox{W.J. Wisniewski\unskip,\iSLAC}
\mbox{J. L. Wittlin\unskip,\iMASS}
\mbox{M. Woods\unskip,\iSLAC}
\mbox{T.R. Wright\unskip,\iWISC}
\mbox{R.K. Yamamoto\unskip,\iMIT}
\mbox{J. Yashima\unskip,\iTOHO}
\mbox{S.J. Yellin\unskip,\iUCSB}
\mbox{C.C. Young\unskip,\iSLAC}
\mbox{H. Yuta\unskip.\iAOMORI}

\it
\vskip \baselineskip                   
\vskip \baselineskip      
\baselineskip=.75\baselineskip   
\iAOMORI
Aomori University, Aomori , 030 Japan, \break
\iBRI
University of Bristol, Bristol, United Kingdom, \break
\iBRUN
Brunel University, Uxbridge, Middlesex, UB8 3PH United Kingdom, \break
\iBU
Boston University, Boston, Massachusetts 02215, \break
\iCOLO
University of Colorado, Boulder, Colorado 80309, \break
\iCSU
Colorado State University, Ft. Collins, Colorado 80523, \break
\iFERR
INFN Sezione di Ferrara and Universita di Ferrara, I-44100 Ferrara, Italy, \break
\iFRAS
INFN Lab. Nazionali di Frascati, I-00044 Frascati, Italy, \break
\iJHU
Johns Hopkins University,  Baltimore, Maryland 21218-2686, \break
\iLBL
Lawrence Berkeley Laboratory, University of California, Berkeley, California 94720, \break
\iMASS
University of Massachusetts, Amherst, Massachusetts 01003, \break
\iMISSI
University of Mississippi, University, Mississippi 38677, \break
\iMIT
Massachusetts Institute of Technology, Cambridge, Massachusetts 02139, \break
\iMOSCOW
Institute of Nuclear Physics, Moscow State University, 119899, Moscow Russia, \break
\iNAGO
Nagoya University, Chikusa-ku, Nagoya, 464 Japan, \break
\iOREG
University of Oregon, Eugene, Oregon 97403, \break
\iOXF
Oxford University, Oxford, OX1 3RH, United Kingdom, \break
\iPERU
INFN Sezione di Perugia and Universita di Perugia, I-06100 Perugia, Italy, \break
\iRAL
Rutherford Appleton Laboratory, Chilton, Didcot, Oxon OX11 0QX United Kingdom, \break
\iRUTG
Rutgers University, Piscataway, New Jersey 08855, \break
\iSLAC
Stanford Linear Accelerator Center, Stanford University, Stanford, California 94309, \break
\iSOONG
Soongsil University, Seoul, Korea 156-743, \break
\iTENN
University of Tennessee, Knoxville, Tennessee 37996, \break
\iTOHO
Tohoku University, Sendai 980, Japan, \break
\iUCSB
University of California at Santa Barbara, Santa Barbara, California 93106, \break
\iUCSC
University of California at Santa Cruz, Santa Cruz, California 95064, \break
\iVAND
Vanderbilt University, Nashville,Tennessee 37235, \break
\iWASH
University of Washington, Seattle, Washington 98105, \break
\iWISC
University of Wisconsin, Madison, Wisconsin 53706, \break
\iYALE
Yale University, New Haven, Connecticut 06511. \break

\rm


\end{center}


\begin{figure}
\vspace*{0.3cm}
\epsfxsize=9.2cm
\epsffile{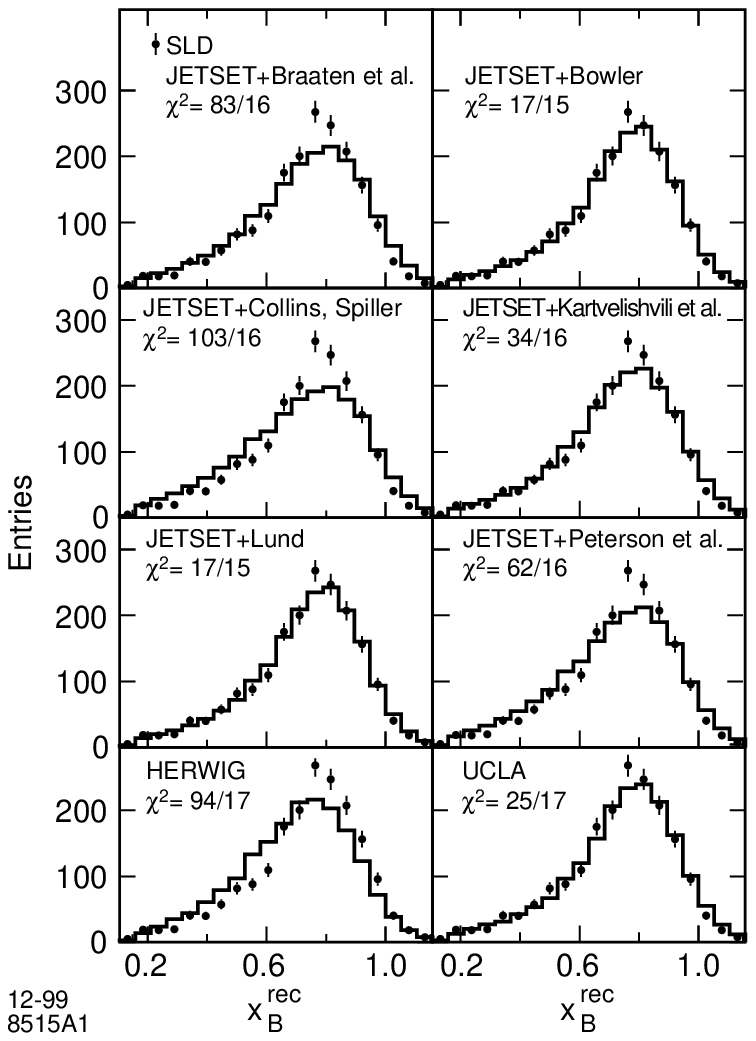}
\vspace*{0.7cm}
\caption{
\label{fig:brec}
Distribution of the reconstructed scaled energy of weakly decaying $B$-hadrons
(points); the errors are statistical.
The predictions of eight models are shown as histograms.
}
\end{figure}

\begin{figure}[h]
\epsfxsize=8.7cm
\epsffile{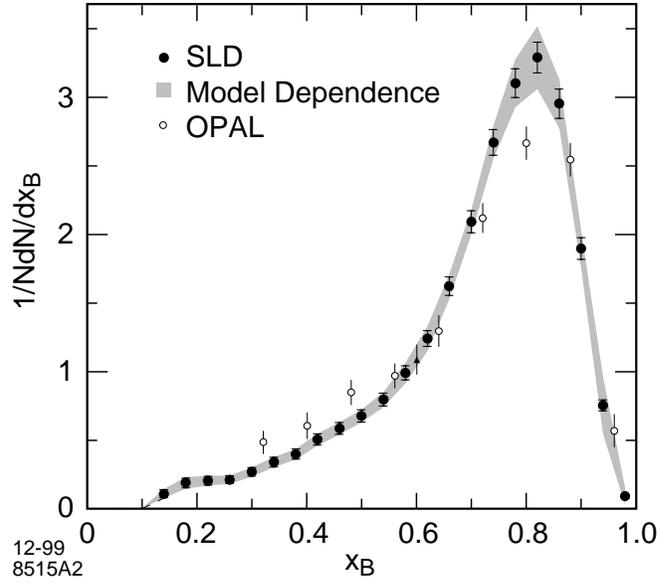}
\vspace*{0.7cm}
\caption{
Unfolded distribution of weakly-decaying scaled $B$-hadron energy (points). The errors 
are statistical only and do not include point-to-point correlations. 
The band represents the envelope of acceptable functions (see text).
Also shown (circles) is the best previous measurement~[15].
}
\label{fig:btrue}
\end{figure}


\begin{thebibliography}{99}
\bibitem{twobs}
In hemispheres with more than one vertex we
considered only the vertex with largest distance from the IP.
\bibitem{negm}
Due to our resolution $M_{0max}^2$ can be negative, 
in which case we set $M_0$ = 0.
\end{thebibliography}
\end{document}